\documentclass{svproc}

\usepackage{url}
\usepackage{graphicx}
\usepackage{verbatim}
\usepackage{tabularx}
\usepackage{hyperref}
\usepackage{booktabs}
\usepackage{cite}
\usepackage{svg}
\usepackage{amsmath}
\usepackage{subcaption}

\begin{document}
\mainmatter

\title{DREAMS: Modelling Support for Research into Engineering and Artistic Design}

\titlerunning{DREAMS: Modelling Tool}

\author{Apala Chakrabarti}

\authorrunning{Apala Chakrabarti}

\institute{Centre of Excellence in Design, IISc\\
\email{chakrabartiapala@outlook.com}}

\maketitle

\begin{abstract}
\par Design Research Methodology (DRM) supports systematic design research through representations such as Reference Models and Impact Models. However, the practical construction and maintenance of these models often remains manual, requiring repeated redrawing, layout adjustment, and separate handling of assumptions, references, and supporting evidence. This can make DRM modelling time-consuming, visually cluttered, and difficult to revise as models increase in complexity. This paper presents DREAMS (Modelling Support for Research into Engineering and Artistic Design), an early-stage prototype modelling environment developed to support the creation and maintenance of DRM Reference Models and Impact Models. The tool enables users to construct typed causal models using DRM-relevant elements, define signed causal relationships, and attach assumptions, experiential inputs, and references directly to causal links. It also provides layout support and search functions to improve readability, modifiability, and retrieval of supporting information. A preliminary comparative evaluation with four DRM users was conducted against manual modelling practice. The results indicate reductions in model creation time, revision time, repositioning effort, edge crossings, and evidence retrieval time when using DREAMS. These findings are interpreted as early evidence of practical potential rather than full validation. The contribution of the paper lies in identifying requirements for DRM-aligned modelling support, presenting the design and implementation of DREAMS, and demonstrating its potential to reduce modelling effort and improve traceability in DRM-based research.

\end{abstract}

\section{Introduction}

Contemporary design research increasingly addresses problems shaped by interacting technical, human, organisational, and contextual factors. Such situations involve interdependencies, feedback effects, and evolving problem framings, making structured externalisation important for analysis, communication, and decision support \cite{1,2}. This is consistent with earlier accounts of design exploration as a co-evolutionary process, where problem understanding and solution development evolve together rather than proceeding in a fixed linear sequence \cite{8}. Within design research, Design Research Methodology (DRM) provides a systematic basis for studying and representing such situations, including through Reference Models (RM) and Impact Models (IM) \cite{1}.

RM and IM support explicit representation of influential factors, causal relationships, assumptions, and outcomes \cite{1}. However, despite their methodological value, their construction remains only weakly supported in practice. Researchers often rely on general-purpose drawing tools or manually arranged diagrams that offer flexibility, but limited support for the semantic requirements of DRM-based modelling. As models grow in size and are revised iteratively, these workflows become increasingly difficult to maintain and reorganise \cite{2}.

The limitation is not simply one of drawing capability. Existing graph visualisation and diagramming tools are primarily intended to organise and display relational structures rather than support DRM-specific modelling semantics \cite{4,5,6,7}. DRM-based models require support for typed elements, signed causal relations, and traceable links between relationships and their supporting assumptions, references, or observations. In most general-purpose tools, such information is difficult to embed directly within the model and is often maintained separately from the graph itself. This results in fragmented workflows and reduces the interpretability and maintainability of the model as it develops.

In response, this paper presents DREAMS (Modelling Support for Research into Engineering and Artistic Design), as an early-stage prototype modelling environment for the creation and maintenance of DRM Reference Models and Impact Models. The name DREAMS is coined from DRM, reflecting the methodology it supports and its application to research in engineering and artistic design. The paper makes three contributions. First, it identifies functional requirements for a DRM-aligned modelling environment. Second, it presents the design rationale and prototype realisation of DREAMS. Third, it reports a preliminary comparative evaluation against manual modelling practice. The contribution should therefore be understood as an initial step towards dedicated support for DRM-based modelling, with early evidence of practical potential rather than full validation.

\section{Background}

\subsection{Requirements of DRM-Based Modelling}

DRM provides a structured basis for representing influencing conditions, design activities, and outcomes in a systematic way \cite{1}. Its application also extends beyond engineering design, including descriptive research into aesthetic design processes \cite{jagtap2015}. In the case of Reference Models and Impact Models, modelling involves more than graphical depiction alone.

The representation must capture directed causal relations, distinguish between different categories of model elements, and retain the assumptions, references, or observations that support those relations. More broadly, systematic design approaches emphasise the need to make design information, dependencies, and rationale explicit so that they can be examined, communicated, and revised during the design process \cite{10}.

Such models are also developed iteratively. As understanding of the problem and its influencing structure evolves, RM and IM representations often need to be extended, revised, and reorganised \cite{2}. Modelling support must therefore address not only initial construction, but also continued modification and maintenance.

\subsection{Graph Visualisation and Readability}

Graph visualisation research has established several layout strategies for improving the readability of node-link representations, including hierarchical, layered, and force-directed approaches \cite{3,4}. Tools such as Graphviz, yEd, and Gephi demonstrate the practical value of automated support for graph arrangement, including spacing, alignment, and crossing reduction in complex structures \cite{5,6,7}.

For causal modelling, layout readability is especially important because interpretation depends on clear tracing of directional relationships. Poorly organised visual representations can also increase cognitive effort, particularly when users must simultaneously interpret model structure, causal direction, and supporting information \cite{9}. Prior work shows that edge crossings and poor structural organisation can significantly hinder graph comprehension \cite{3}. Readable layout is therefore a necessary requirement for tools intended to support RM and IM construction.

\subsection{Limitations of Existing Tool Support}

Despite these strengths, existing graph visualisation tools remain primarily general-purpose systems. They support graph construction and layout, but offer limited support for the domain-specific semantics required in DRM-based modelling. In particular, they do not adequately integrate typed factors, signed causal links, and relationship-level traceability to assumptions, references, or observations within a single modelling environment.

In practice, this leads to fragmented workflows in which the graphical model is maintained separately from the supporting rationale and evidence associated with it. As models increase in scale and complexity, this separation makes them harder to interpret, revise, and retrieve information from. The central gap, therefore, is not the absence of graph drawing capability, but the lack of an integrated modelling environment aligned with the semantic and operational requirements of DRM.

\section{Methodology}

\subsection{Requirements Derivation}

Functional requirements were derived from three sources: the limitations identified in the preceding literature review, the representational demands of DRM-based modelling reported in the literature, and a short survey involving 16 researchers using DRM in their research. Participants were identified through DRM-related research networks and workshops and represented varying levels of experience: seven were first-time or recent DRM users, five had used DRM for approximately two years, and four had used it throughout their doctoral research. The survey examined current modelling practices and the main practical difficulties encountered during RM and IM construction.

Respondents were asked about the time typically spent creating DRM-based models, whether these models were produced by hand or using software tools, the approximate size of the models they worked with, and the main challenges encountered during model creation, revision, and maintenance. Additional questions addressed how supporting assumptions, references, and other evidential information were recorded and revisited during the modelling process. The survey was therefore used both to identify recurring difficulties and to assess the practical need for more integrated tool support.

\begin{table}[!t]
\centering
\caption{Summary of recurring issues reported in the DRM user survey ($n=16$)}
\label{tab:survey_summary}
\resizebox{0.94\linewidth}{!}{
\begin{tabular}{p{7.2cm} c}
\hline
\textbf{Reported issue} & \textbf{Respondents reporting issue} \\
\hline
Repeated manual restructuring during model revision & 100\% \\
Difficulty maintaining layout readability as models grew & 68.75\% \\
Supporting assumptions/references maintained separately from the model & 93.75\% \\
Difficulty retrieving assumptions, references, or specific links & 100\% \\
Limited support for typed elements and causal polarity in existing tools & 100\% \\
\hline
\end{tabular}}
\end{table}

Across the literature review and survey responses, four issues emerged consistently. First, DRM models often required repeated manual restructuring as they increased in size and link density. Second, existing tools provided limited support for DRM-specific causal semantics, including typed model elements and signed relationships. Third, supporting assumptions, references, and observations were frequently maintained separately from the model itself, reducing traceability. Fourth, retrieval of specific model content became increasingly difficult as diagrams grew in complexity. Overall, the combined findings indicated a clear practical need for a modelling environment that better supports structured, traceable, and modifiable DRM representations.

Based on this combined analysis, six requirements were defined for the prototype. These represent the core functional capabilities needed for a DRM-aligned modelling environment and are summarised in Table~\ref{tab:requirements}.

\begin{table}[!t]
\centering
\caption{Requirements for a DRM-aligned modelling tool}
\label{tab:requirements}
\resizebox{0.94\linewidth}{!}{
\begin{tabular}{l p{10.8cm}}
\hline
\textbf{Requirement ID} & \textbf{Requirement description} \\
\hline
R1 & Support representation of causal relationships with explicit directionality and polarity between model elements \\
R2 & Allow definition and differentiation of typed model elements, such as influencing factors, success factors, key factors, and assumptions \\
R3 & Integrate supporting information, including assumptions, references, and other contextual evidence, directly within the model structure \\
R4 & Provide layout assistance to improve readability and reduce the effort required for manual restructuring \\
R5 & Support iterative modification of the model without requiring complete redrawing or loss of representational consistency \\
R6 & Enable search and retrieval of model elements, relationships, and associated supporting information \\
\hline
\end{tabular}}
\end{table}

Taken together, these requirements reflect both the representational semantics of DRM and the practical needs reported by researchers actively using DRM in their work. For the preliminary evaluation, the assessment was narrowed to those aspects most directly observable in a comparative modelling task: layout readability, structural clarity, traceability, and modifiability effort. These aspects were selected because they correspond to the most recurring practical difficulties identified in the survey and literature review, while also reflecting the central purpose of DREAMS as a modelling support environment.

\subsection{Comparative Assessment Criteria}

To compare the prototype with current DRM modelling practice, four assessment criteria were selected from the broader set of requirements and practical difficulties identified earlier. The comparison was conducted against manual modelling because the survey indicated that DRM Reference and Impact Models are typically constructed by hand or using general-purpose drawing tools, rather than specialised causal-loop or systems-modelling software. Manual modelling was therefore selected as the baseline most representative of current practice.

Layout readability concerns the visible extent of edge crossings, node overlap, and local visual congestion. Structural clarity concerns how clearly directional causal paths and source-target relationships can be identified and followed. Traceability concerns the presence and accessibility of links between model relationships and their associated assumptions, references, or observations. Modifiability effort concerns the extent of manual repositioning, redrawing, or restructuring required after model changes.

These criteria are summarised in Table~\ref{tab:evaluation_criteria}.

\begin{table}[!t]
\centering
\caption{Criteria for comparative assessment of modelling approaches}
\label{tab:evaluation_criteria}
\resizebox{0.96\linewidth}{!}{
\begin{tabular}{l p{4.2cm} p{7.4cm}}
\hline
\textbf{Criterion ID} & \textbf{Criterion} & \textbf{Assessment basis} \\
\hline
C1 & Layout readability & Visible extent of edge crossings, node overlap, and local visual congestion in the resulting model \\
C2 & Structural clarity & Clarity with which directional causal paths and source-target relationships can be identified and followed \\
C3 & Traceability & Presence and accessibility of links between relationships and associated assumptions, references, or other supporting information \\
C4 & Modifiability effort & Extent of manual repositioning, redrawing, or restructuring required to create and maintain a readable and consistent model \\
\hline
\end{tabular}}
\end{table}

Together, these criteria provide the basis for examining whether DREAMS addresses the practical limitations of manual DRM modelling in a more coherent and usable manner.

\section{Prototype Realisation}

Based on the identified requirements, DREAMS was realised as a prototype modelling environment for constructing and revising DRM-based Reference Models (RM) and Impact Models (IM). The prototype was designed to provide an integrated workspace in which the core representational and practical demands of DRM-based modelling could be addressed within a single environment.

To support DRM-specific causal representation, signed causal relationships were used to represent both direction and polarity of influence, while typed nodes were used to distinguish between relevant model element categories. Supporting information, including assumptions and references, was embedded within the model structure in order to maintain traceability between represented relationships and their evidential basis.

For layout support, a layered or hierarchical arrangement was adopted because RM and IM are directional and causal in character, and this form of organisation supports readability in node-link representations \cite{4,3}. In addition, the prototype employed an editable graph structure to support iterative revision without complete redrawing, together with text-based search to enable retrieval of model elements, relationships, and associated supporting information.

\begin{figure}[!t]
\centering
\includegraphics[
    width=0.88\linewidth,
    trim=10 10 10 10,
    clip
]{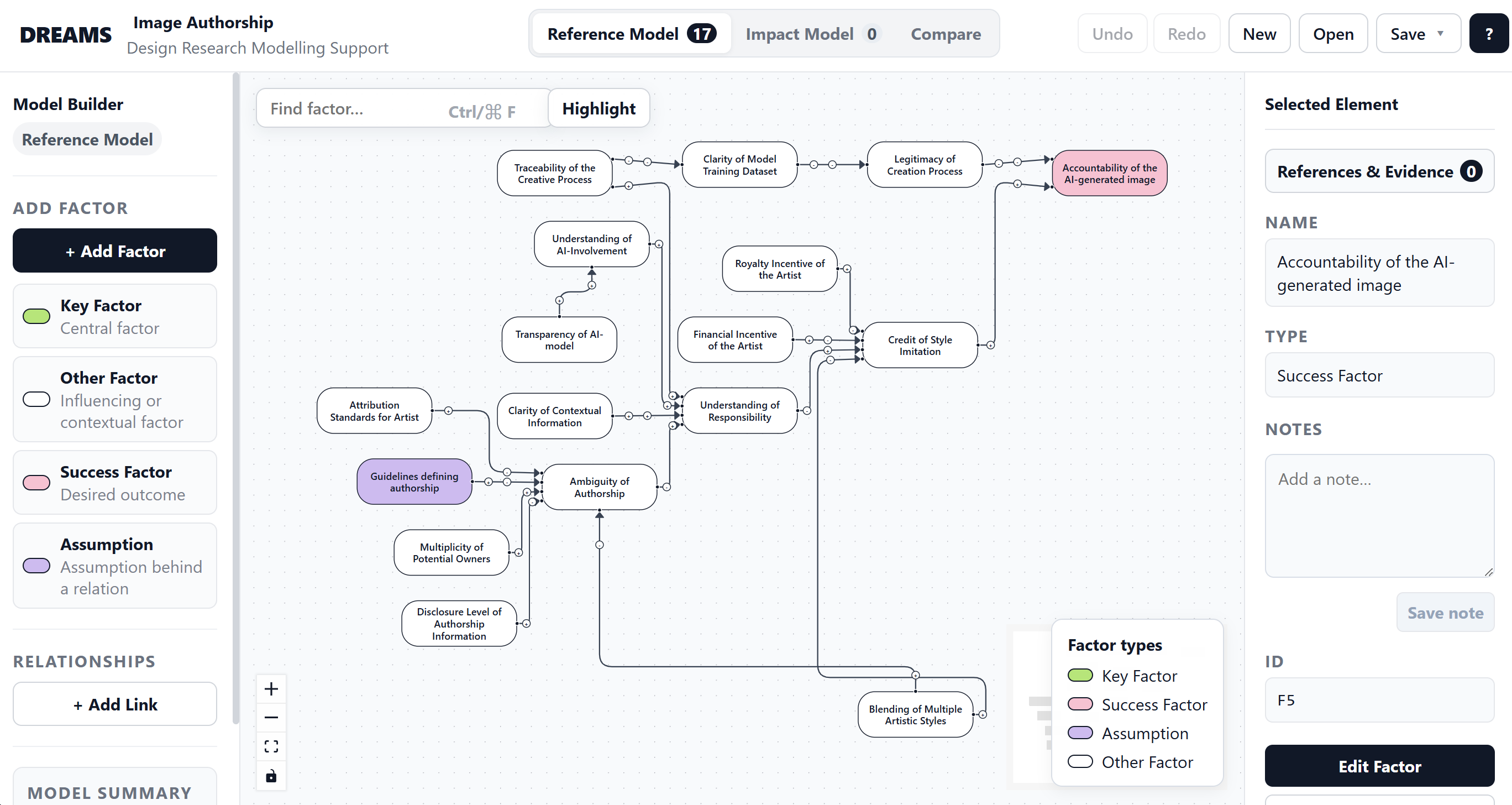}
\caption{User interface of the DREAMS prototype.}
\label{fig:dreams_ui}
\end{figure}

Figure~\ref{fig:dreams_ui} shows the user interface of the DREAMS prototype. The interface provides a unified modelling workspace in which users can construct DRM-based models, define typed elements, specify signed causal links, attach supporting information, and iteratively refine the representation. In this way, the prototype brings together model construction, modification, and traceability within a single interactive environment.

A working version of DREAMS is publicly available at
\url{https://dreams-black.vercel.app}. The prototype is provided to enable researchers to explore the modelling environment, construct DRM Reference and Impact Models, and evaluate its use in their own research contexts.

\section{Preliminary Evaluation}

A preliminary comparative evaluation was conducted between manual modelling and prototype-supported modelling using DREAMS. Four second- and third-year PhD researchers, familiar with DRM through prior training using the methodology described by Blessing and Chakrabarti [1], participated in the evaluation. No prior training in DREAMS was provided. All participants completed the same model creation, revision, and evidence-retrieval tasks under both conditions. To reduce order and learning effects, the condition order was counterbalanced, with two participants completing manual modelling first and two completing DREAMS first. The comparison focused on differences in representation, organisation, revision, and retrieval rather than on differences in model content.

\begin{figure}[!t]
\centering

\begin{subfigure}[c]{0.40\linewidth}
    \centering
    \includegraphics[
        width=\linewidth,
        height=4.5cm,
        keepaspectratio
    ]{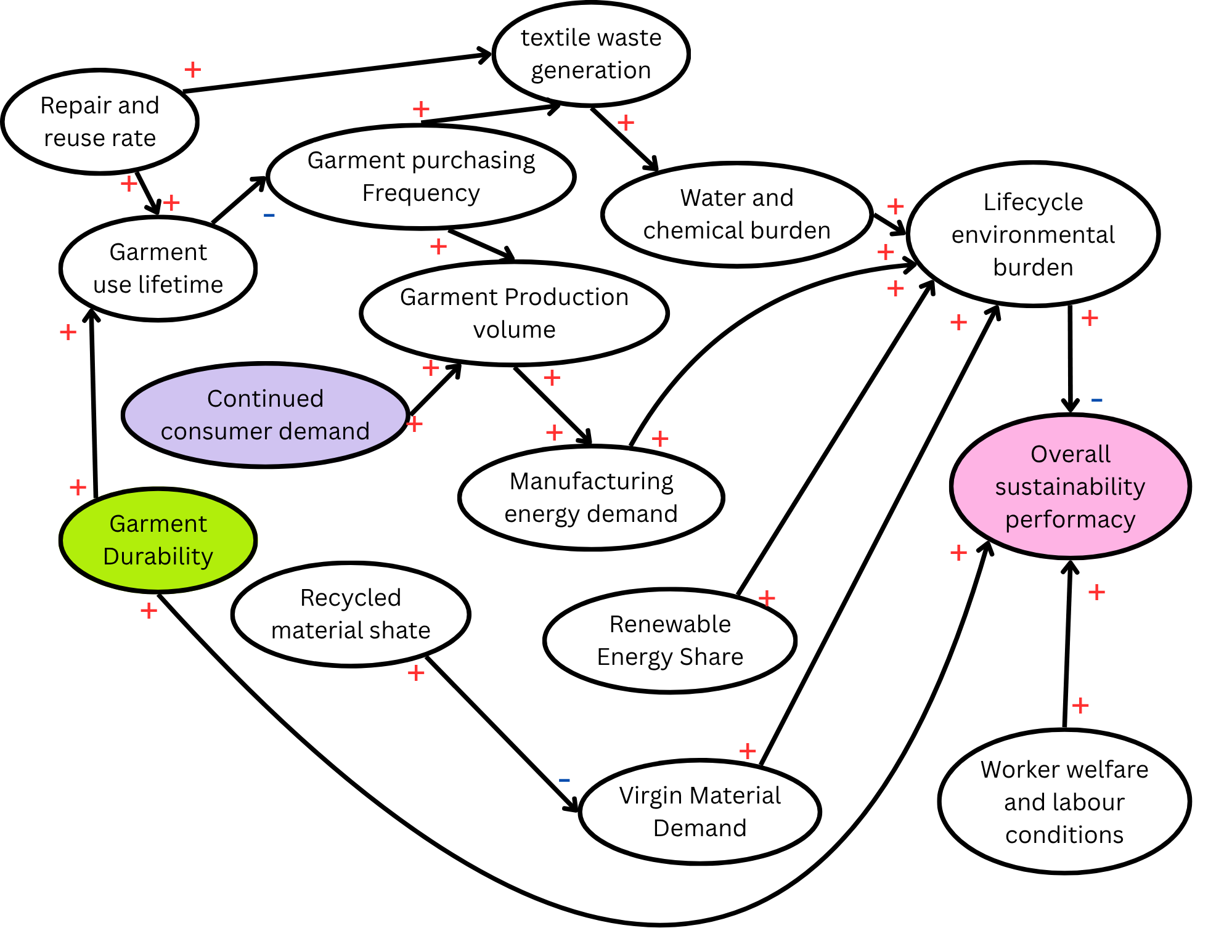}
    \caption{Manual modelling approach}
    \label{fig:manual_model}
\end{subfigure}
\hfill
\begin{subfigure}[c]{0.56\linewidth}
    \centering
    \includegraphics[
        width=\linewidth,
        height=4.5cm,
        keepaspectratio
    ]{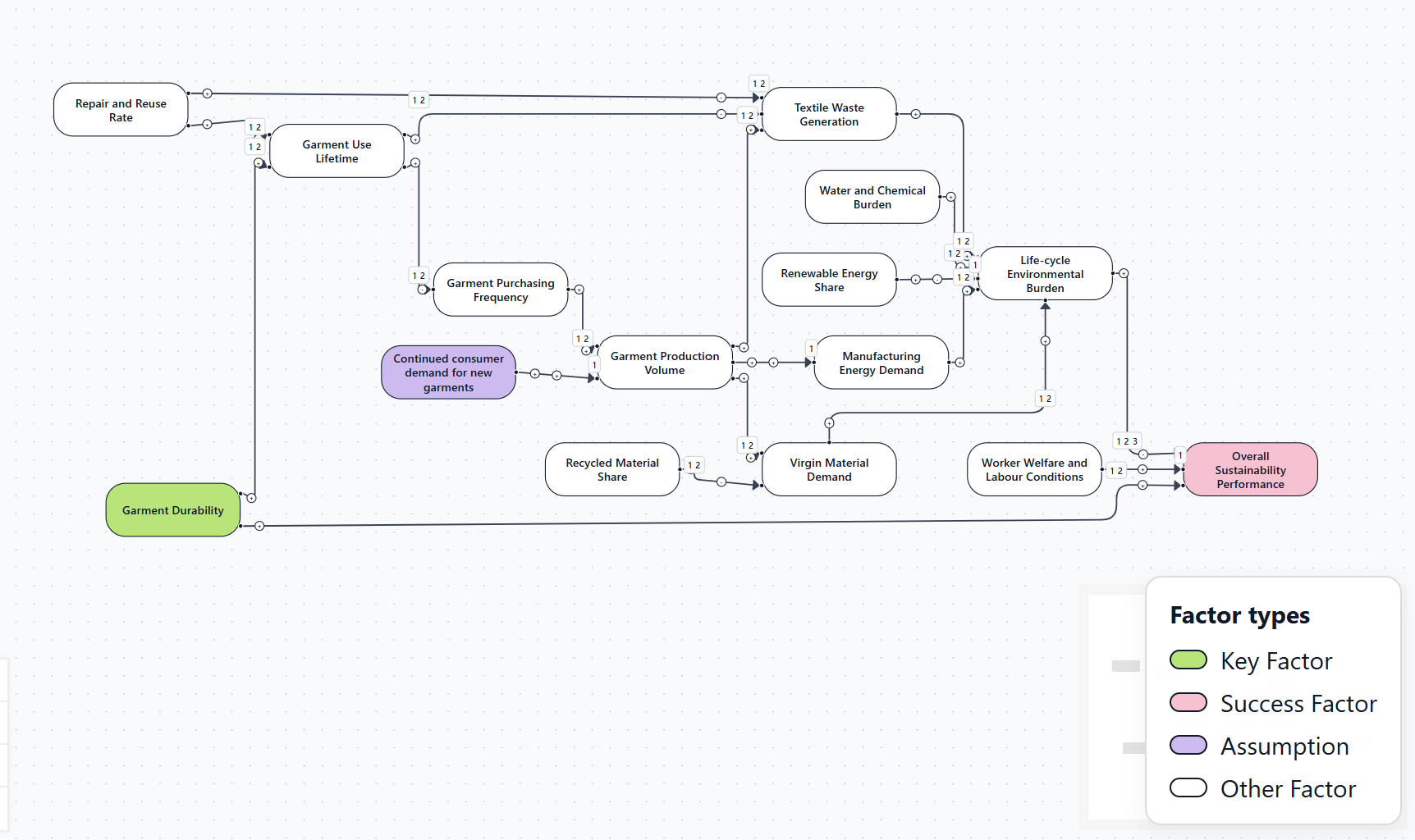}
    \caption{Equivalent model in DREAMS}
    \label{fig:dreams_model}
\end{subfigure}

\caption{Comparison between manual DRM modelling and the DREAMS prototype participant models.}
\label{fig:manual_vs_dreams}
\end{figure}

Participants were provided with the same sample Reference Model data. For each condition, initial model creation was timed without revision, after which participants performed the same prescribed revision task. The evaluation was based on the four assessment criteria defined earlier: layout readability, structural clarity, traceability, and modifiability effort. These criteria were examined through five observable measures: total model creation time, revision time, number of edge crossings, number of repositioning actions, and evidence retrieval time. For the retrieval task, participants were asked to locate a specified supporting reference. All models under both conditions were created entirely by the participants without intervention from the tool developer. Figure~\ref{fig:manual_vs_dreams} shows representative participant-generated models under the two conditions.

The purpose of the evaluation was to examine whether the prototype addressed the practical limitations identified earlier more effectively than manual modelling practice. Given the small number of participants, the findings are interpreted as an initial indication of usefulness rather than as complete validation.

\subsection{Comparative Findings}

Table~\ref{tab:comparative_results} summarises the quantitative comparison across the four participants. Time-based measures were recorded during task completion, repositioning actions from the observed editing process, and edge crossings from the resulting models.

\begin{table}[!t]
\centering
\caption{Comparative evaluation of manual modelling and DREAMS ($n=4$)}
\label{tab:comparative_results}
\resizebox{0.98\linewidth}{!}{
\begin{tabular}{p{3.8cm} c c c c}
\hline
\textbf{Measure} & \textbf{Unit} & \textbf{Manual (mean)} &
\textbf{DREAMS (mean)} & \textbf{Reduction} \\
\hline
Model creation time & min & 51.0 & 22.0 & 56.9\% \\
Revision time & min & 24.5 & 2.0 & 91.8\% \\
Repositioning actions & count & 37.5 & 0.0 & 100.0\% \\
Edge crossings & count & 4.25 & 1.0 & 76.5\% \\
Evidence retrieval time & min & 5.0 & 1.0 & 80.0\% \\
\hline
\end{tabular}}
\end{table}

\begin{figure}[!t]
\centering

\begin{subfigure}[t]{0.47\linewidth}
    \centering
    \includegraphics[width=\linewidth]{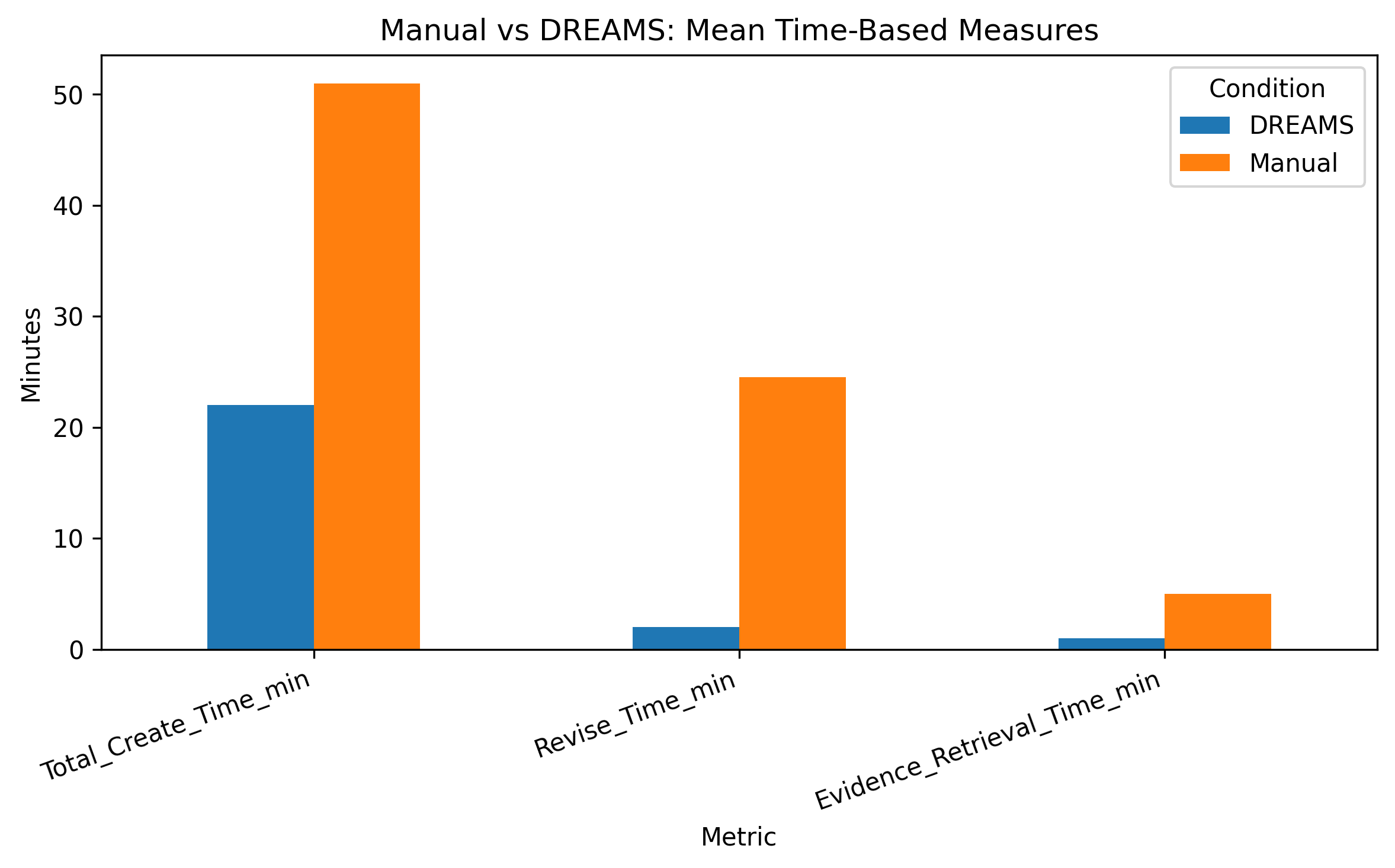}
    \caption{Time-based measures}
    \label{fig:time_metrics}
\end{subfigure}
\hfill
\begin{subfigure}[t]{0.47\linewidth}
    \centering
    \includegraphics[width=\linewidth]{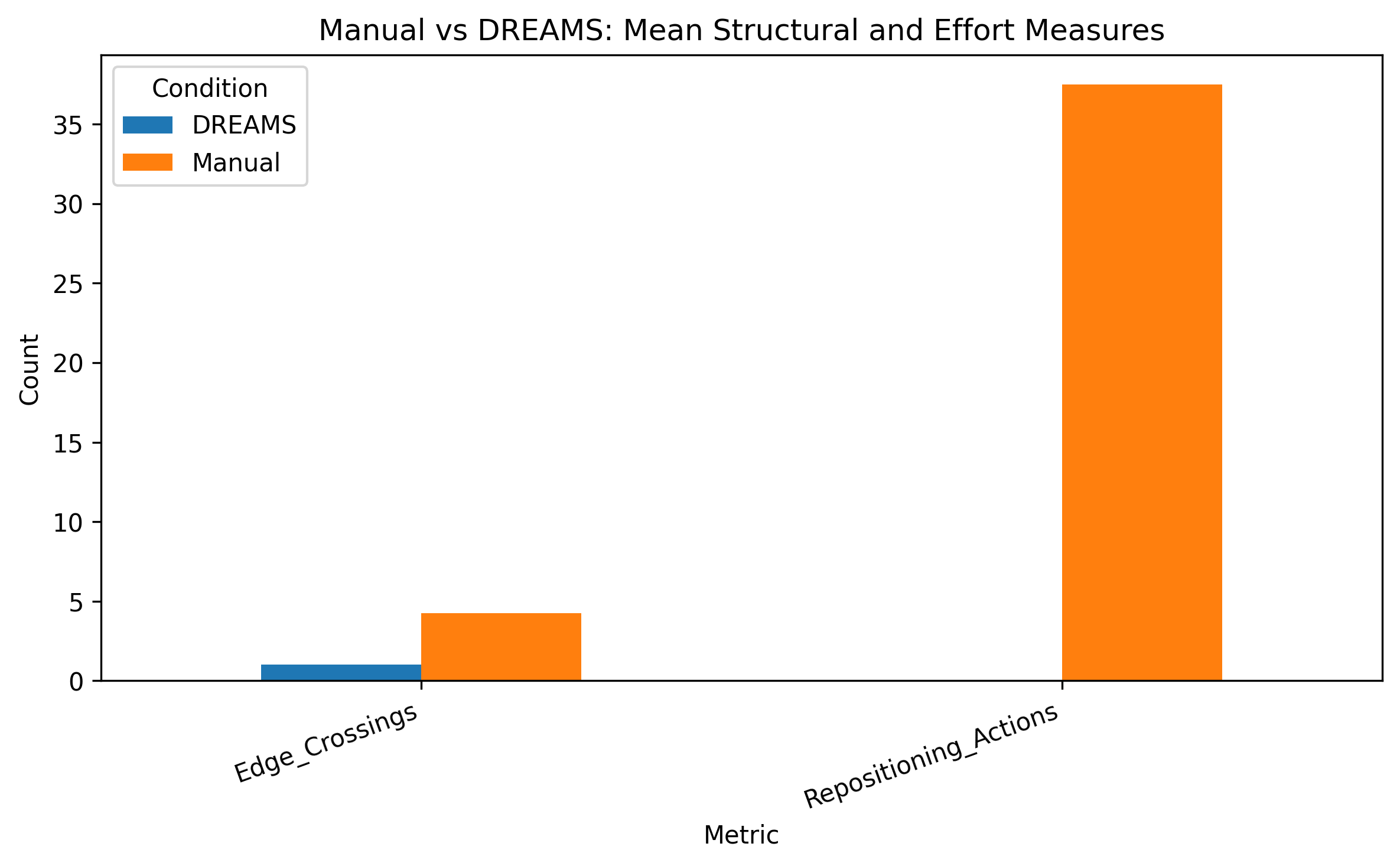}
    \caption{Structural and effort-related measures}
    \label{fig:structure_metrics}
\end{subfigure}

\caption{Comparison of mean performance measures for manual modelling and DREAMS across four participants.}
\label{fig:quantitative_comparison}
\end{figure}

Figure~\ref{fig:quantitative_comparison} summarises the differences in time-based and structure-related measures.

The largest differences were observed in revision effort, repositioning actions, and evidence retrieval. Manual modelling required substantial local adjustment after changes, whereas DREAMS reduced revision time from a mean of 24.5 minutes to 2.0 minutes and eliminated repositioning actions in the recorded tasks. Evidence retrieval time was also reduced from a mean of 5.0 minutes in the manual condition to 1.0 minute in DREAMS, reflecting the integration of supporting information within the model structure.

Differences were also observed in model creation time and layout readability. Mean creation time was reduced from 51.0 minutes to 22.0 minutes, while the mean number of visible edge crossings decreased from 4.25 to 1.0. These results indicate that the prototype provided a more structured basis for constructing and revising DRM-based models while preserving clearer relationships between model elements and their associated supporting information.

\section{Discussion}

The results indicate that the main value of DREAMS lies in providing modelling support aligned with the semantic and operational requirements of DRM rather than in graph drawing alone. General-purpose tools can produce workable diagrams, particularly for smaller models, but they rely on manual arrangement and external handling of supporting information. As model complexity increases, this makes readability, traceability, and revision more difficult to maintain.

The evaluation should be interpreted in relation to the broader requirements identified earlier in the paper. While six functional requirements were defined for the prototype, the preliminary evaluation focused on four assessment criteria that were most directly observable within a comparative modelling task: layout readability, structural clarity, traceability, and modifiability effort. The five reported measures therefore do not constitute a separate set of criteria, but serve as observable indicators of these four assessment dimensions. This distinction is important because the present study evaluates the practical modelling support provided by DREAMS, rather than attempting to validate every functional requirement exhaustively.

The evaluation also indicates that layout support alone is insufficient for DRM-based modelling. Readability depends not only on spatial organisation, but also on semantic clarity. In DREAMS, typed elements, signed causal relations, and embedded supporting information together supported a representation that was easier to interpret, revise, and retrieve information from than the manually produced version.

At the same time, the contribution of DREAMS should be understood in a limited and specific sense. The prototype does not automate causal reasoning or remove the interpretive judgement required in RM and IM construction. Its contribution lies in providing a modelling environment that better supports the representation, modification, and traceability requirements of DRM. The findings therefore indicate that DRM-based modelling can benefit from more dedicated support than is currently available in general-purpose graph and diagramming tools.

\section{Limitations and Future Work}

This study is limited by the preliminary nature of the evaluation and the early-stage status of the prototype. The evaluation involved four participants and a single modelling case, and therefore does not establish statistical generalisability. In addition, the study did not examine collaborative use, larger models, or comparison with specialised causal-loop or systems-modelling software. Given the small sample size, the quantitative results are presented descriptively and are not intended to support statistical inference.

The current prototype also has technical limitations. The layered layout strategy supports directional organisation, but it does not guarantee optimal readability in all cases, particularly as model size and link density increase. Similarly, while the prototype improves traceability by linking supporting information to relationships, the underlying causal structure remains dependent on user judgement and is not automatically validated.

Future work should therefore extend the evaluation across more users and modelling cases, incorporate broader quantitative measures, and refine the interaction and layout capabilities of the prototype. Further development should also address collaborative use, larger-scale model management, and more advanced retrieval and filtering functions.



\begin{thebibliography}{20}

\bibitem{1}
Blessing, L. T. M., \& Chakrabarti, A. (2009).
\textit{DRM: A Design Research Methodology}.
Springer.
\url{https://doi.org/10.1007/978-1-84882-587-1}

\bibitem{2}
Dorst, K., \& Cross, N. (2001).
Creativity in the design process: Co-evolution of problem-solution.
\textit{Design Studies, 22}(5), 425--437.
\url{https://doi.org/10.1016/S0142-694X(01)00009-6}

\bibitem{8}
Maher, M. L., Poon, J., \& Boulanger, S. (1996).
Modeling design exploration as co-evolution.
\textit{Microcomputers in Civil Engineering, 11}(3), 195--209.
\url{https://doi.org/10.1111/j.1467-8667.1996.tb00323.x}

\bibitem{jagtap2015}
Jagtap, S., \& Jagtap, S. (2015).
Aesthetic design process: Descriptive design research and ways forward.
In A. Chakrabarti (Ed.), \textit{ICoRD'15 -- Research into Design Across Boundaries, Vol. 1}
(pp. 375--385).
Springer India.
\url{https://doi.org/10.1007/978-81-322-2232-3_33}

\bibitem{4}
Gansner, E. R., Koutsofios, E., North, S. C., \& Vo, K.-P. (1993).
A technique for drawing directed graphs.
\textit{IEEE Transactions on Software Engineering, 19}(3), 214--230.
\url{https://doi.org/10.1109/32.221135}

\bibitem{5}
Ellson, J., Gansner, E., Koutsofios, E., North, S. C., \& Woodhull, G. (2002).
Graphviz---Open source graph drawing tools.
In \textit{Graph Drawing} (pp. 483--484).
Springer.
\url{https://doi.org/10.1007/3-540-45848-4_57}

\bibitem{6}
Bastian, M., Heymann, S., \& Jacomy, M. (2009).
Gephi: An open source software for exploring and manipulating networks.
In \textit{Proceedings of the International AAAI Conference on Web and Social Media, 3}(1), 361--362.
\url{https://gephi.org/publications/gephi-bastian-feb09.pdf}

\bibitem{7}
yWorks GmbH (2023).
\textit{yEd Graph Editor}.
\url{https://www.yworks.com/products/yed}

\bibitem{10}
Pahl, G., Beitz, W., Feldhusen, J., \& Grote, K.-H. (2007).
\textit{Engineering Design: A Systematic Approach} (3rd ed.).
Springer.
\url{https://doi.org/10.1007/978-1-84628-319-2}

\bibitem{3}
Purchase, H. C. (1997).
Which aesthetic has the greatest effect on human understanding?
In G. Di Battista (Ed.), \textit{Graph Drawing (GD 1997)} (pp. 248--261).
Springer.
\url{https://doi.org/10.1007/3-540-63938-1_67}

\bibitem{9}
Sweller, J., Ayres, P., \& Kalyuga, S. (2011).
\textit{Cognitive Load Theory}.
Springer.
\url{https://doi.org/10.1007/978-1-4419-8126-4}

\end{thebibliography}
\end{document}